\documentstyle[epsf,seceq,twocolumn]{jpsj}

\author{Akihisa {\sc Koga},
Seiya {\sc Kumada}, 
and Norio {\sc Kawakami}}
\title{
Two-Dimensional Quantum Spin Systems with Ladder 
and Plaquette Structures
}
\inst{Department of Applied Physics, Osaka University, 
Suita, Osaka 565-0871}
\recdate{September 14, 1998}
\abst{
We investigate low-energy properties of two-dimensional 
quantum spin systems with the ladder and  plaquette structures,
which are described by a generalized antiferromagnetic Heisenberg model 
with  both of the bond and spin alternations.
By exploiting a non-linear $\sigma$ model  technique 
and a modified spin wave approach, we 
evaluate the spin gap and the spontaneous magnetization to
discuss the quantum phase transition between the ordered and 
disordered states. We argue how the spin-gapped phase is driven to 
the antiferromagnetic phase in the phase diagram.
}
\kword{2D spin system, spin gap, non-linear $\sigma$ model,
 modified spin wave}
\def\runtitle{2D Quantum Spin Systems with Ladder and 
Plaquette Structure}
\def\runauthor{Akihisa {\sc Koga}, 
Seiya {\sc Kumada} and Norio {\sc Kawakami}}
\begin{document}
\sloppy
\maketitle
\section{Introduction}
Recent intensive studies on high-$T_{c}$ superconductors
and related materials 
have attracted renewed interest in the quantum phase transitions
in low-dimensional spin systems. 
In particular, a variety of new materials have been 
found experimentally: For instance, CaV$_{4}$O$_{9}$
\cite{cavao1,cavao2,cavao3}, 
which may be described by the two-dimensional (2D) 
Heisenberg model on a depleted square lattice,
realizes a quantum spin liquid even in two dimension. 
Also,  SrCu$_{2}$O$_{3}$\cite{srcuo1,srcuo2}
has been intensively studied
as a prototypical example of ladder compounds,
\cite{exnmag,nonmag} 
for which the introduction of nonmagnetic impurities 
induces a novel phase transition to 
the magnetic state.  For these materials, the
plaquette or ladder structure is essential to 
stabilize the non-magnetic phase with spin gap.
The spin chains  with  an alternating array
of different spins (mixed spin chains or alternating spin chains)
also give another new paradigm,
\cite{AltExp,AltThe,AltExa,TonHik,Fukui} for which 
the topology of spin arrangement 
plays an essential role to determine the 
low-energy properties. Quantum phase transitions in the above 
various low-dimensional spin systems have provided an 
interesting subject in quantum spin systems.

In our previous paper,\cite{previous} 
by employing non-linear $\sigma$ model (NL$\sigma$M) techniques,
we investigated the competition  
between the gapful and gapless states in 
a generalized  spin ladder which possesses
both of the spin and bond alternations.
We systematically investigated not only 
ladder systems but also mixed spin chains and
plaquette-type spin chains, and  clarified how 
the competing interactions in addition to topological properties 
control the quantum phase transitions. 
In real materials, however, the coupling between ladders 
or chains neglected in our previous paper becomes certainly 
important, and  
may drive the system to the ordered state.

Motivated by the above points, in this paper we 
study the quantum phase transitions between 
the ordered and disordered states in 2D spin systems,
by extending the previous NL$\sigma$M analysis of the spin chain 
and ladder systems.
In particular, we deal with the 2D antiferromagnetic 
Heisenberg model with the ladder  
and plaquette structures, which also possesses 
the spatial variation of spins (mixed spins).
By calculating the spin gap and the spontaneous magnetization
by means of NL$\sigma$M and modified spin wave (MSW) approaches,
\cite{TakahashiAF,HirschTang}
we discuss how the disordered state
with spin gap competes against the magnetically ordered state. 

The paper is organized as follows. In $\S$2 we introduce 
a generalized 2D spin model for which  
both of the bond and spin alternations are included.
In $\S$3, by extending the method of S\'en\'echal,\cite{Senechal}
we calculate the spin gap at finite temperatures by
means of NL$\sigma$M with saddle-point approximation, and then
argue to what extent this approach can describe the 
quantum  phase transitions for 2D systems.
We also point out the shortcoming in this approach for 
spin-$1/2$ systems. In $\S$ 4, we further study the model by 
using the MSW approach, which qualitatively improves the  
results of NL$\sigma$M in the spin-$1/2$ case.
Brief summary is given in $\S$5.

%
%
\section{Model Hamiltonian}
\begin{figure}[htb]
\vspace{0.2cm}
\epsfxsize=7cm 
\centerline{\epsfbox{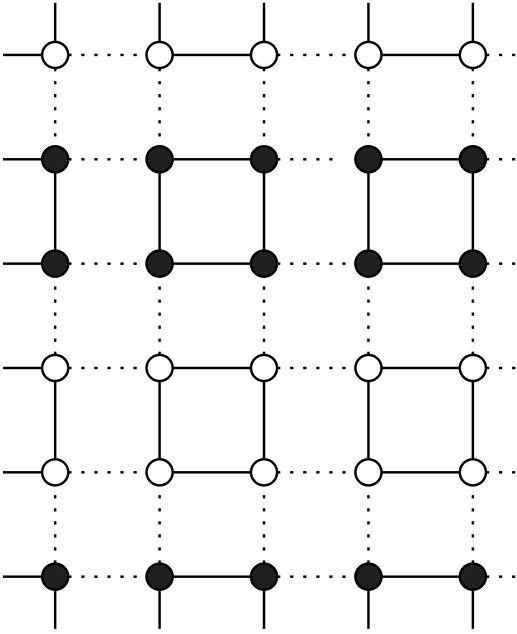}} 
\caption{A generalized 2D spin system 
with both of the bond and spin alternations.}
\label{fig:Model}
\end{figure}
We introduce a 2D antiferromagnetic 
spin system with both of the bond  and spin alternations, 
which is described by the following Hamiltonian
on a square lattice,
\begin{eqnarray}
H&=&J_{x}\sum_{i,j}\left[1+(-1)^{i}\gamma_{x}\right]{\mib S}_{i,j}\cdot
{\mib S}_{i+1,j}\nonumber\\
&+&J_{y}\sum_{i,j}\left[1+(-1)^{j}\gamma_{y}\right]{\mib S}_{i,j}\cdot
{\mib S}_{i,j+1}\label{eqn:H},
\end{eqnarray}
where $J_{x}$ $(J_{y})$ and $\gamma_{x}$ $(\gamma_{y})$ are the 
exchange coupling and the bond-alternation parameter 
along the $x$ $(y)$ direction, respectively. 
Here ${\mib S}_{i,j}$ is the spin operator 
at the $(i,j)$-th site in the $x-y$ plane. 
Since we deal with the antiferromagnetic case with $J_{x}, J_{y}>0$ and 
$-1\leq\gamma_{x},\gamma_{y}\leq 1$, the ground state of the 
Hamiltonian (\ref{eqn:H}) becomes the N\'eel state
in the classical limit. What is distinct from an
ordinary 2D square lattice model is that the present system 
includes not only the bond alternation but also the 
spin alternation. 
For simplicity, we here consider a specific arrangement  of 
two types of spins shown schematically  in Fig. \ref{fig:Model}, 
where black circles and white circles denote the 
spins $s_{1}$ and $s_{2}$. 
The generalized Hamiltonian (\ref{eqn:H})  allows us to study 
a variety of interesting spin systems which possess a ladder structure,
a plaquette structure, and a mixture of different spins.
This is a natural extension of our previous analysis of 
the mixed-spin chains and ladders \cite{previous}
to more realistic cases
including the effects of two dimensionality.
In the next section we shall first study the model by employing 
the NL$\sigma$M techniques.

%
%
\section{Non-Linear Sigma Model Analysis}
In this section we discuss the competition between
the spin-gapped phase and the magnetically 
ordered phase  by taking the non-linear $\sigma$ model
(NL$\sigma$M) as a low energy effective theory.
Let us begin by briefly summarizing 
how  NL$\sigma$M techniques are applied to our model in the 
continuum limit.
Using the coherent-state path integral formalism,\cite{stand}
the partition function in the present system is given by
\begin{eqnarray}
Z=\int {\rm D}\mib\Omega_{ij}\exp\left[ -{\rm i}\sum_{ij}
s_{ij}\omega\left[\mib\Omega_{ij}\right]-\int_{0}^{\beta}
{\rm d}\tau H\left(\tau\right)\right],\nonumber\\
\label{eqn:Zinit}
\end{eqnarray}
where 
\begin{eqnarray}
\omega\left[{\mib\Omega}_{ij}\right]&=&\int_{0}^{\beta}{\rm d}\tau
\dot{\varphi}_{ij}(1-\cos\theta_{ij}),\\
H(\tau )&=&J_{x}\sum_{i,j}s_{i,j}s_{i+1,j}\left[ 1+(-1)^{i}\gamma_{x}\right]
{\mib\Omega}_{i,j}\cdot{\mib\Omega}_{i+1,j}\nonumber\\
&+&J_{y}\sum_{i,j}s_{i,j}s_{i,j+1}\left[1+(-1)^{j}\gamma_{y}\right]
{\mib\Omega}_{i,j}
\cdot{\mib\Omega}_{i,j+1}.\nonumber\\\label{eqn:Htau}
\end{eqnarray}
In the above expressions we have introduced 
the unit vector ${\mib\Omega}$ by ${\mib S}=
s{\mib \Omega}$, where ${\mib\Omega}=(\sin\theta\cos\varphi
,\sin\theta\sin\varphi,\cos\theta)$. The term $\omega\left[
{\mib\Omega}\right]$ is the Berry phase.
In the present approach we assume that 
the usual spin-wave analysis can correctly describe low-energy modes 
at wave vectors near $(0, 0), (0, \pi), (\pi, 0)$, and $(\pi, \pi)$. 
The last one corresponds to the ordering wave vector.
Then the unit vector ${\mib\Omega}_{i,j}$ can be written in 
terms of three fluctuation fields
and one staggered field as follows\cite{Senechal} 
\begin{equation}
\mib\Omega_{2i+\alpha,2j+\beta}=(-1)^{\alpha+\beta}\mib n
\left(\mib r'\right),\label{eqn:omega}
\end{equation}
\begin{equation}
\mib n\left(\mib r'\right)=\phi\left(\mib r\right)+
a{\sum_{mn}}^{'}(-1)^{m\alpha+n\beta}\mib l_{mn}^{(k)}\left(\mib r\right),
\end{equation}
where $\mib r'=\mib r+\alpha m+\beta n$, $m(n)$ 
and $\alpha(\beta)$ runs from $0$ to $1$, and $a$ is the 
lattice constant. The primed sum means that 
the term $m=n=0$ is omitted.
Therefore fluctuations which we take 
here are $l_{01}$, $l_{10}$, and $l_{11}$ for each site.
The point we wish to notice is that 
the fluctuation $\mib{l}$ is further
distinguished by the index $k(=1,2)$,
corresponding to spins $s_{1}$ and $s_{2}$, respectively. 
The staggered field $\mib\phi$ is 
a slowly varying field on the scale of the lattice spacing. 
Substituting eq.(\ref{eqn:omega}) 
into eq.(\ref{eqn:Htau}) and making the expansion up to 
quadratic order in $\mib l, \mib\phi'$, and $\dot{\mib\phi}$, 
we arrive at  the following Hamiltonian in the continuum limit
\begin{eqnarray}
H\left(\tau\right)&=&\int {\rm d}\mib r \left(h_{x}+h_{y}\right),\\
h_{x}&=&\frac{J_{x}}{2}\left[
\left( s_{1}^{2}+s_{2}^{2}\right)\left(1+\gamma_{x}\right)
\left(\partial_{x}\mib\phi\right)^{2}\right.\nonumber\\
&&+2\left(1+\gamma_{x}\right)\partial_{x}\mib\phi\cdot\left(s_{1}^{2}
\mib l_{10}^{(1)}+s_{2}^{2}\mib l_{10}^{(2)}\right)\nonumber\\
&&+2\sum_{j}\left.\left(s_{1}^{2}{\mib l_{1j}^{(1)}}^{2}+
s_{2}^{2}{\mib l_{1j}^{(2)}}^{2}\right)\right],\\
h_{y}&=&\frac{J_{y}}{4}\left[ 2\left(1-\gamma_{y}\right)
\sum_{i}\left(s_{1}^{2}{\mib l_{i1}^{(1)}}^{2}+
s_{2}^{2}{\mib l_{i1}^{(2)}}^{2}\right)\right.\nonumber\\
&&+4s_{1}s_{2}\left(1+\gamma_{y}\right)
\left[\left(\partial_{y}\mib\phi\right)^{2}+
\partial_{y}\mib\phi\cdot\left(\mib l_{01}^{(1)}+
\mib l_{01}^{(2)}\right)\right]\nonumber\\
&&\left.+s_{1}s_{2}\left(1+\gamma_{y}\right)
\sum_{i}\left(\mib l_{i1}^{(1)}+\mib l_{i1}^{(2)}\right)^{2}\right].
\end{eqnarray}
The Berry phase in this system is given by 
\begin{eqnarray}
S_{\rm Berry}&=&{\rm i}\sum_{ij}s_{ij}\omega
\left[\mib\Omega_{ij}\right]\nonumber\\
&=&\frac{\rm i}{2a}\int {\rm d}\mib r\left(s_{1}\mib l_{11}^{(1)}+
s_{2}\mib l_{11}^{(2)}\right)\cdot\mib\phi\times\dot{\mib\phi}.
\end{eqnarray}
Integrating eq. (\ref{eqn:Zinit}) over the 
fluctuation fields ${\mib l}$, we thus end up with the NL$\sigma$M
\begin{eqnarray}
Z&=&\int{\rm D}{\mib\phi}\exp\left
\{-\frac{1}{2g}\left[\frac{1}{v}\left(
\partial_{\tau}{\mib\phi}\right)^{2}\right.\right.\nonumber\\
&&\left.\left.+vR_{x}\left(\partial_{x}{\mib\phi}\right)^{2}
+vR_{y}\left(\partial_{y}{\mib\phi}\right)^{2}
\right]\right\}.\label{eqn:NLSM}
\label{gappu}
\end{eqnarray}
Note that the Berry phase terms cancel out
in the present system and do not show up in the 
effective action. This simple result 
allows us to further investigate the quantum phase transitions
by employing ordinary NL$\sigma$M techniques.
Since general forms of parameters $g, v, R_{x}$, 
and $R_{y}$ are rather complicated, we shall show their concrete 
formulae for each case discussed in the following.

\subsection{Gap equation}

To proceed the calculation based on the NL$\sigma$M, 
further approximations may be needed. 
Here we use the saddle point approximation 
to obtain the spin gap equation from eq.(\ref{gappu}), 
by extending the treatment of  S\'en\'echal.\cite{Senechal}
The validity of this approximation will be discussed 
later in this section. 
In what follows, we set the speed  $v=1$; it can 
be restored by dimensional analysis at the end of the calculation.
Rescaling the field $\mib\phi\rightarrow\sqrt{g}\mib\phi$, 
the Lagrangian (\ref{eqn:NLSM}) is now  written as
\begin{eqnarray}
L&=&\frac{1}{2}\left[\left(\partial_{\tau}{\mib\phi}\right)^{2}+R_{x}
\left(\partial_{x}{\mib\phi}\right)^{2}+
R_{y}\left(\partial_{y}{\mib\phi}
\right)^{2}\right.\nonumber\\
&&\left.-\sigma\left({\mib\phi}^{2}-1/g\right)\right],
\label{eqn:Lag1}
\end{eqnarray}
where $\sigma$ is the Lagrange multiplier to enforce the constraint
 $\mib\phi^{2}=1/g$. 
Integrating eq.(\ref{eqn:Lag1}) over the field $\mib\phi$, 
we obtain the effective potential to the field $\sigma$
\begin{eqnarray}
V(\sigma )&=&\frac{\sigma}{2g}-\frac{1}{\beta}\sum_{\omega_{n}}
\int\frac{{\rm d}\mib k}{(2\pi )^{2}}\nonumber\\
&&\times\log\left( 1+\frac{\sigma}{\omega_{n}^{2}+R_{x}k_{x}^{2}+
R_{y}k_{y}^{2}}\right).
\label{pote}
\end{eqnarray}
Here we replace the field $\sigma$ by the uniform value, 
which corresponds to the square of the spin gap.
To evaluate the spin gap, the magnon speed $v$ is restored by 
dimensional analysis.
Finding out the saddle point of the potential eq.(\ref{pote}), 
we obtain the spin gap equation
\begin{eqnarray}
\frac{1}{g}=\int&&\frac{{\rm d}\mib k}{(2\pi )^{2}}\frac{1}{
\sqrt{R_{x}k_{x}^{2}+R_{y}k_{y}^{2}+(\Delta /v)^{2}}}\nonumber\\
&&\times\coth\frac{\beta}{2}\sqrt{R_{x}k_{x}^{2}+R_{y}
k_{y}^{2}+(\Delta /v)^{2}}.
\label{eq:GAP}
\end{eqnarray}
It is to be noticed that 
a momentum cut-off parameter $\Lambda$,
which is proportional to the inverse lattice 
spacing $a^{-1}$, should be introduced in
the above expression. 
Since it is hard to determine the cut-off procedure
only from the present theory, we use other means to fix it.
We outline a procedure for the case of plaquette 
structure characterized by the  parameters: 
$\gamma=\gamma_{x}=\gamma_{y}, J=J_{x}=J_{y}$. 
In the limit $\gamma\rightarrow 1$ the system changes to 
the assembly of plaquette singlets. Then the spin gap of the 
system is exactly $\Delta=2J$ at $T=0$.
Requiring that the gap obtained from eq.(\ref{eq:GAP}) 
with $\gamma=1$ should be equal to $2J$,
the cut-off parameter is chosen as $a\Lambda=\sqrt{2\pi/3}$.
Assuming that this value can be used for 
other values of  $\gamma$, we discuss the low-energy properties 
of spin systems with plaquette structure.

This completes our formulation of the model 
in terms of NL$\sigma$M approach.
In the following we show the numerical results 
obtained for the 2D spin systems with plaquette 
and  ladder structures,  as well as the mixed spin system. 

\subsection{Plaquette-structure system}

Let us start our discussions for 
the  spin system with plaquette structure
with  the  parameters, $\gamma=\gamma_{x}=\gamma_{y}, J=J_{x}=J_{y}$, 
for which spins  are assumed to be uniform ($s=s_{1}=s_{2}$). 
In this case, the parameters in NL$\sigma$M  are given by
\begin{equation}
\begin{array}{rcl}
R_{x}&=&\left( 1-\gamma_{x}^{2}\right)j,\\
R_{y}&=&1-\gamma_{y}^{2},\\
g&=&2a\sqrt{1+j}/s,\\
v&=&2J_{y}sa\sqrt{1+j},
\end{array}
\label{eqn:param1}
\end{equation}
where $j$ is defined by $j=J_{x}/J_{y}$. 
The cut-off parameter, as already mentioned, 
is taken as $a\Lambda=\sqrt{2\pi /3}$.

\begin{figure}[htb]
\vspace{0.1cm}
\epsfxsize=8cm 
\centerline{\epsfbox{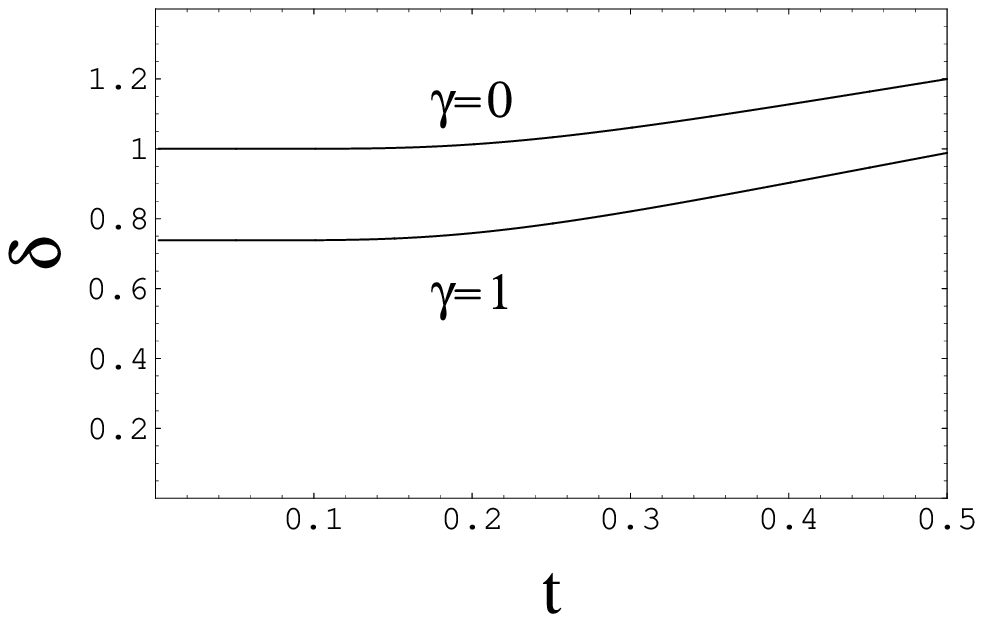}} 
\epsfxsize=8cm 
\centerline{\epsfbox{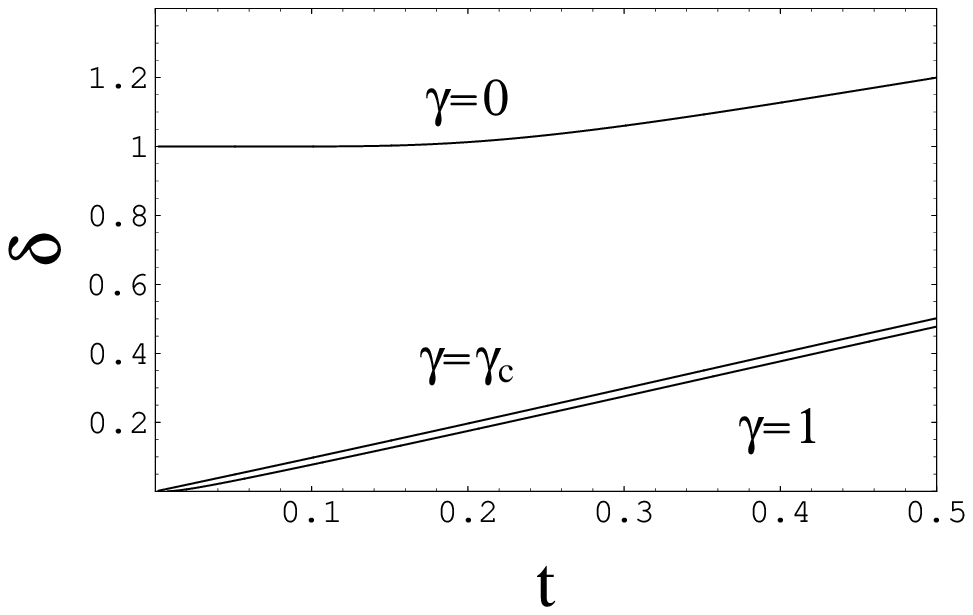}} 
\caption{Normalized spin gap $\delta(=\Delta /\Delta_{0})$ 
for spin $s=1/2$ (a) and $s=1$ (b) plaquette structures calculated 
by the NL$\sigma$M. It is shown as a function $t(=aT)$.
$\Delta_{0}$ is the spin gap at $T=0$.
}
\label{fig:nplagap}
\end{figure}
The behavior of the normalized spin gap
$\delta(=\Delta /\Delta_{0})$ is shown as a function 
of the temperature $t(=aT)$ for spin 
$s=1/2$ (Figs. \ref{fig:nplagap}(a)) 
and $s=1$ (Figs. \ref{fig:nplagap}(b)) systems.
$\Delta_{0}$ is the spin gap at $T=0$.
Physically, the spin gap should be regarded as the 
inverse of the correlation length.
As the temperature is decreased, 
the spin gap (inverse correlation length) decreases,
and reaches zero or a finite value according to the 
nature of ground state.
For the $s=1$ case, we can clearly observe the quantum phase
transition at $T=0$ in  Fig. \ref{fig:nplagap}(b)
when we decrease the value of $\gamma$ from unity.
At $\gamma=1$, the system at $T=0$ is composed of isolated 
plaquettes and forms the plaquette
singlet ground state which  possesses the spin gap
in the excitation.
In decreasing $\gamma$, the plaquette-singlet
state with spin gap is driven to 
the magnetically ordered state, which
is realized for $\gamma \leq \gamma_{c}$.
As is the case for ordinary 2D quantum systems
with continuous symmetry,
the ordered state is allowed only for $T=0$, which  
is correctly described in our treatment, as is seen from
the figure. Although the NL$\sigma$M approach 
has produced qualitatively correct results for the case
with $s>1/2$, we can not find the 
quantum phase transition to the ordered state for $s=1/2$, 
as seen from Fig. \ref{fig:nplagap} (a).
This is contradicted to the  
well-known fact that the ground state of the $s=1/2$ isotropic square 
lattice ($\gamma=0$) should be an antiferromagnetically ordered state.
This implies that our approach may not  
sufficiently take into account antiferromagnetic
fluctuations, and this shortcoming may come
mainly from the saddle point
approximation we used. In order to 
improve our results for the  $s=1/2$ case, we further study 
the system by employing modified spin wave approach in 
the next section.

\subsection{Ladder-structure system}
Next we investigate the 2D
system  with ladder structure which is realized by 
choosing $\gamma_{x}=0$ and $\gamma=\gamma_{y}$.
This is the system composed of periodic array of ladders 
along $y$-direction.\cite{KatoImada}
By appropriately tuning the interaction strength, 
we can naturally interpolate from the isolated 
ladders to the 2D system with ladder structure. 
The corresponding couplings for NL$\sigma$M is given by 
the same expression as (\ref{eqn:param1}).
\begin{figure}[htb]
\vspace{0.1cm}   
\epsfxsize=8cm 
\centerline{\epsfbox{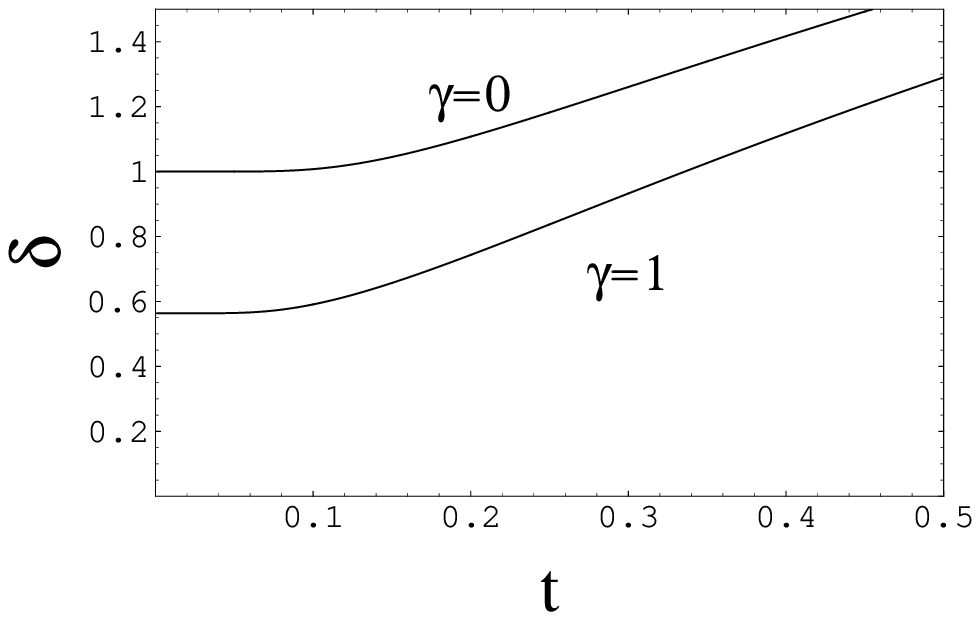}} 
\epsfxsize=8cm 
\centerline{\epsfbox{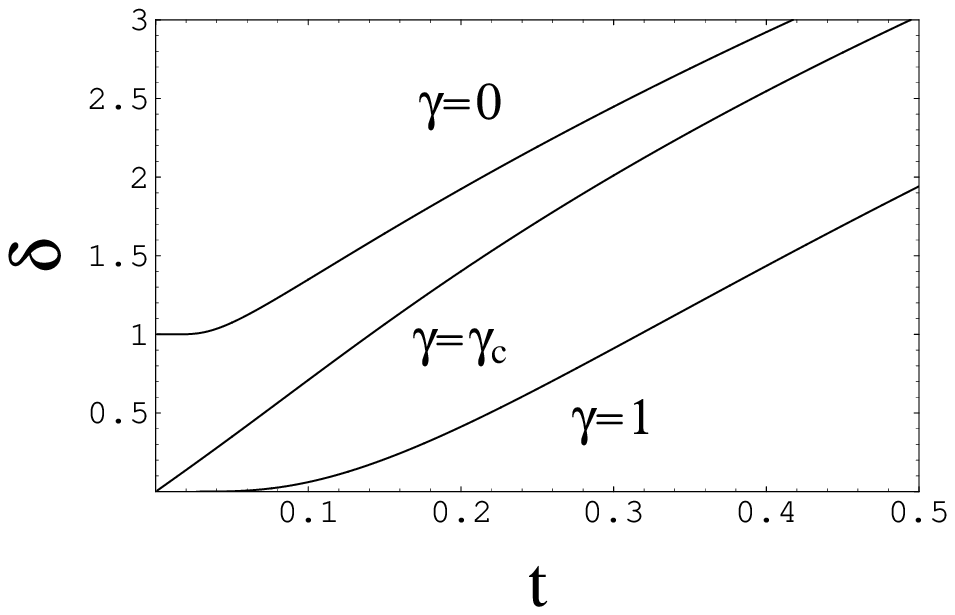}} 
\caption{Normalized spin gap 
$\delta(=\Delta /\Delta_{0})$ for the spin $s=1/2$ (a) and $s=1$ 
(b) ladder systems calculated by the NL$\sigma$M ($j=1/2)$. It 
is shown as a function $t(=aT)$. }
\label{fig:nladgap}
\end{figure}
The cut-off parameter in the ladder system is taken as $a\Lambda=0.99$,
which  is determined according to the numerical results\cite{Dagotto,White} 
($\Delta/J=0.504$) for the uniform isotropic ladder ($j=1/2$). 
The results for the spin gap are shown in  Figs. \ref{fig:nladgap}
for the cases of $s=1/2$ and $1$. Note that 
the system with the parameter $\gamma=0 (1)$ corresponds 
to the isotropic square lattice (the isolated ladder).
For the isolated ladder, the system should have the spin 
gap, which is  consistent with our results.
By decreasing $\gamma$ from unity,
we observe the quantum phase transition at $T=0$ from the 
spin-gapped phase to the magnetic phase for $s=1$, but
not for $s=1/2$. As is the case for the 
model with plaquette structure,  we thus describe the 
quantum phase transition properly for $s>1/2$.
The problem we have encountered again for $s=1/2$ 
has the same origin mentioned before for the 
plaquette case.
We shall come to this problem in the next section. 

\subsection{Mixed spin system}
So far, we have considered the uniform-spin systems.
We note that mixed spin chains with periodic array of several 
kind of different spins have also attracted much attention both
experimentally and theoretically.
\cite{AltExp,AltThe,AltExa,TonHik} 
For example,  
mixed spin chains with either the ferrimagnetic ground state
or the singlet ground state have been intensively studied. 
\cite{AltThe,AltExa,TonHik} 
In the previous paper,\cite{previous} we have studied 
the mixed spin chains by NL$\sigma$M approach.
To make the theoretical model
more realistic, it may be natural to take into account
the coupling among mixed chains. We address 
this problem in this section. As a simple example 
of mixed spin systems, 
we  here consider the spin system composed of two 
different spins $s_{1}=1$ and $s_{2}=3/2$ (and also 
$s_1=1/2$ and $s_2=1$) for the 2D system shown in Fig. \ref{fig:Model}. 
This type of system enables us to bridge 2D quantum
systems and mixed spin chains mentioned above.
To make our discussions simpler,
we deal with the system without bond alternation. 
Then parameters in  NL$\sigma$M for the mixed spin system 
are given by 
\begin{eqnarray}
R_{x}&=&\frac{j}{8}\frac{\left(1+\alpha\right)^{2}
\left(1+\alpha^{2}\right)}{\alpha^{2}},\\
R_{y}&=&1,\\
g&=&\frac{1+\alpha}{s_{2}}aK,\\
v&=&\frac{4J_{y}s_{2}a}{1+\alpha}K,
\end{eqnarray}
where $K=\sqrt{(2j+1)
[(1+\alpha)^{2}+4j\alpha]/[(1+\alpha)^{2}+8j\alpha]}$
and $\alpha=s_{2}/s_{1}$. 
\begin{figure}[htb]
\vspace{0.1cm}
\epsfxsize=8cm 
\centerline{\epsfbox{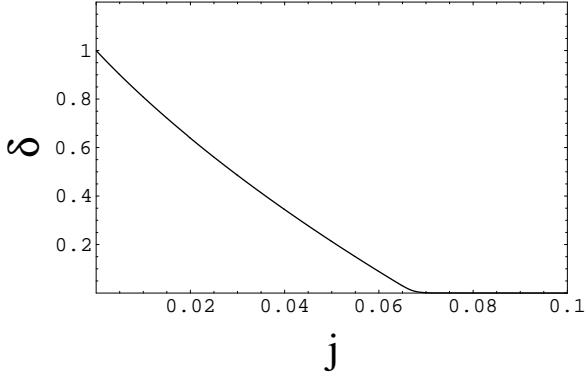}} 
\caption{Normalized spin gap $\delta(=\Delta/\Delta_{0})$ 
for the mixed spin system, for $s_1=1$, $s_2=3/2$, 
calculated by the NL$\sigma$M. 
It is shown as a function $j(=J_{x}/J_{y})$.}
\label{fig:naltgap}
\end{figure}
The cut-off parameter is taken
as $a\Lambda=1.41$ according to 
the numerical results in ref\cite{Night}: 
$\Delta /J=0.41\; (T=0, s_{1}=s_{2}=1, j=0)$.
The normalized spin gap calculated 
for $s_1=1$ and $s_2=3/2$ 
is shown in Fig. \ref{fig:naltgap}.
For $j=0$, the system is composed of isolated mixed spin chains,
which is known to be massive, being
consistent with the present results.\cite{previous}
When the value of $j$ is increased, the system gradually
changes to the 2D system. We then find a plausible result
that there exists the critical value of $j$ at which
the system is changed to the ordered state.
For the $s_1=1/2$ and $s_2=1$ case , however, 
we have checked that in the whole range of $j$ the system 
has the spin gap, and the long range order does not show up.
Since we naturally expect that for moderately large $j$
the antiferromagnetic ground  state may be stabilized,
we should improve our treatment for this case,
which is reexamined in the next section by
MSW approach.


%
%
\section{Modified Spin-Wave Approach}

We have so far studied our 2D spin model by the NL$\sigma$M with 
saddle point approximation, by employing the method
of S\'en\'echal \cite{Senechal},
which is a natural extension of our previous study 
on the spin chain and spin ladder systems.\cite{previous}
Although qualitative features 
for the quantum phase transitions have been described 
for the systems with larger spins, we have encountered the
problem for the $s=1/2$ cases as well as the mixed spin case
with $s=1/2, 1$, for which our NL$\sigma$M analysis may not
describe the correct phase transition.
We think that this shortcoming comes mainly
>from our saddle point approximation, which
may not properly incorporate magnetic
correlations for $s=1/2$ case.
To improve this point, there may be several possible
ways. The one is to apply renormalization-group analysis 
to the NL$\sigma$M instead of saddle point approximation,
which was successfully applied to 2D $s=1/2$ systems.
\cite{CHN}
Alternatively, there exists a different but similar
approach based on the modified spin wave (MSW)
approximation.\cite{TakahashiAF,HirschTang}
 This method is  known to provide  
rather satisfactory results for the isotropic $s=1/2$ 
spin systems, which are comparable to
the renormalization group method mentioned above.
For instance, both of the above methods have 
produced the correct temperature dependence of the
correlation length for the 2D $s=1/2$ spin systems on 
square lattice. Therefore, the MSW approach
is expected to provide a  sensible rout to
improve our results for small $s$ cases in the 
previous section. In what follows,
we treat the present model within the MSW approximation.
We show the formulation 
in the case of uniform spins, for simplicity,
since it is straightforward to extend it to 
the system composed of two distinct spins.\par
 
By the Holstein-Primakoff transformation, 
the spin operators are 
expressed in terms of boson creation and annihilation operators 
as follows
\begin{equation}
\begin{array}{ll}
\begin{array}{l}
S_{\mibs r}^{z}=s-a_{\mibs r}^{\dag}a_{\mibs r}\\
S_{\mibs r}^{+}=\sqrt{2s-
a_{\mibs r}^{\dag}a_{\mibs r}}\;\;a_{\mibs r}
\end{array}
&{\mib r}=(2i,2j)\in A,\\
\begin{array}{l}
S_{\mibs r}^{z}=b_{\mibs r}^{\dag}b_{\mibs r}-s\\
S_{\mibs r}^{+}=b_{\mibs r}^{\dag}\sqrt{2s-
b_{\mibs r}^{\dag}b_{\mibs r}}
\end{array}
&{\mib r}=(2i+1,2j)\in B,\\
\begin{array}{l}
S_{\mibs r}^{z}=c_{\mibs r}^{\dag}c_{\mibs r}-s\\
S_{\mibs r}^{+}=c_{\mibs r}^{\dag}\sqrt{2s-
c_{\mibs r}^{\dag}c_{\mibs r}}
\end{array}
&{\mib r}=(2i,2j+1)\in C,\\
\begin{array}{l}
S_{\mibs r}^{z}=s-d_{\mibs r}^{\dag}d_{\mibs r}\\
S_{\mibs r}^{+}=\sqrt{2s-
d_{\mibs r}^{\dag}d_{\mibs r}}\;\;d_{\mibs r}
\end{array}
&{\mib r}=(2i+1,2j+1)\in D.\\
\end{array}
\end{equation}
Boson operators $a_{\mibs r}, b_{\mibs r}, c_{\mibs r}, d_{\mibs r}$
belong to the A, B, C, D sublattices, respectively.
Since the unit cell in our system includes four sites, 
we have introduced four kinds of Bose operators. 
In case that we consider the system with two kinds of spins,
 eight kinds of bosons should be introduced.
The Fourier transformations are defined as
\begin{eqnarray}
\alpha_{\mibs r}=\sqrt{\frac{4}{N}}\sum_{\mibs k}\alpha_{\mibs k}
e^{-{\rm i}{\mibs k}\cdot{\mibs r}}&,&
\beta_{\mibs r}=\sqrt{\frac{4}{N}}\sum_{\mibs k}\beta_{\mibs k}
e^{{\rm i}{\mibs k}\cdot{\mibs r}},
\end{eqnarray}
for $\alpha=a,d (\beta=b,c)$, where $N$ is the total number of lattice, 
and the summation over momenta $\mib k$ is restricted to the quarter 
of the original Brillouin zone. 
Then Hamiltonian becomes
\begin{eqnarray}
H&=&2J_{y}s\sum_{\mibs k}\left\{ (1+j)\left(a_{\mibs k}^{\dag}
a_{\mibs k}+b_{\mibs k}^{\dag}b_{\mibs k}+c_{\mibs k}^{\dag}c_{\mibs k}+
d_{\mibs k}^{\dag}d_{\mibs k}\right)\right.\nonumber\\ \nonumber
&&+j\kappa_{x}\left[\left(a_{\mibs k}b_{\mibs k}+
c_{\mibs k}^{\dag}d_{\mibs k}^{\dag}\right) e^{{\rm i}\phi_{x}}+
{\rm H. c. }\right]\\ 
&&\left.+\kappa_{y}\left[\left(a_{\mibs k}c_{\mibs k}+
b_{\mibs k}^{\dag}d_{\mibs k}^{\dag}\right) e^{{\rm i}\phi_{y}}+
{\rm H. c. }\right]\right\}
\label{eqn:HBB},
\end{eqnarray}
where 
\begin{eqnarray}
\kappa_{x}&=&\sqrt{1-(1-\gamma_{x}^{2})\sin^{2}k_{x}},\nonumber\\
\phi_{x}&=&\arg\left[\cos k_{x}+{\rm i}
\gamma_{x}\sin k_{x}\right],\nonumber\\
\kappa_{y}&=&\sqrt{1-(1-\gamma_{y}^{2})\sin^{2}k_{y}},\nonumber\\
\phi_{y}&=&\arg\left[\cos k_{y}+{\rm i}\gamma_{y}
\sin k_{y}\right].\nonumber
\end{eqnarray}
The diagonalization of the Hamiltonian (\ref{eqn:HBB}) yields 
the conventional spin wave excitations.
The staggered magnetization per site is expressed as
\begin{eqnarray}
m&=&\frac{1}{N}\left[\sum_{{\mibs r}\in A}S_{\mibs r}^{z}-
\sum_{{\mibs r}\in B}S_{\mibs r}^{z}-\sum_{{\mibs r}\in C}
S_{\mibs r}^{z}+
\sum_{{\mibs r}\in D}S_{\mibs r}^{z}\right]\nonumber\\
&=&s-\frac{1}{N}\sum_{\mibs k}\left[a_{\mibs k}^{\dag}a_{\mibs k}+
b_{\mibs k}^{\dag}b_{\mibs k}+c_{\mibs k}^{\dag}c_{\mibs k}+
d_{\mibs k}^{\dag}d_{\mibs k}\right].\label{eqn:stg}
\end{eqnarray}
Here it should be noticed that the sublattice 
symmetry \cite{HirschTang} in the ground state is broken.
In order to extend our study into the region of 
the disordered phase, we restore this symmetry by 
enforcing the following constraint on eq.(\ref{eqn:HBB}),
\begin{eqnarray}
s=\frac{1}{N}\sum_{\mibs k}\left[a_{\mibs k}^{\dag}a_{\mibs k}+
b_{\mibs k}^{\dag}b_{\mibs k}+c_{\mibs k}^{\dag}c_{\mibs k}+
d_{\mibs k}^{\dag}d_{\mibs k}\right].\label{eqn:con}
\end{eqnarray}
This means that the staggered 
magnetization (\ref{eqn:stg}) should be zero in the disordered phase.
Since the constraint may give rise to the spin gap
in a certain parameter region, we can discuss the phase transition 
>from the ordered phase to the disordered phase with spin gap.
Using the Lagrange multiplier $\lambda$, the Hamiltonian with 
the constraint (\ref{eqn:con}) is cast to
\begin{eqnarray}
H^{'}=H-\lambda\sum_{\mibs k}\left[a_{\mibs k}^{\dag}a_{\mibs k}+
b_{\mibs k}^{\dag}b_{\mibs k}+c_{\mibs k}^{\dag}c_{\mibs k}+
d_{\mibs k}^{\dag}d_{\mibs k}\right].
\label{hamihami}
\end{eqnarray}
By the Bogoliubov transformations
\begin{equation}
\left(
\begin{array}{c}
a_{\mibs k}\\b_{\mibs k}^{\dag}\\c_{\mibs k}^{\dag}\\d_{\mibs k}
\end{array}
\right)=\frac{1}{\sqrt{2}}\left(
\begin{array}{cc}
U_{\mibs k}^{+}&U_{\mibs k}^{-}\\
V_{\mibs k}^{+}&V_{\mibs k}^{-}
\end{array} 
\right)\left(
\begin{array}{l}
\alpha_{\mibs k}^{(1)}\\\alpha_{\mibs k}^{(2)
\dag}\\\beta_{\mibs k}^{(1)}\\
\beta_{\mibs k}^{(2)\dag}
\end{array}
\right),
\end{equation}
where 
\begin{equation}
U_{\mibs k}^{\pm}=\mp e^{-{\rm i}\phi_{y}}\left(
\begin{array}{cc}
-\cosh\theta_{\mibs k}^{\pm}&
\sinh\theta_{\mibs k}^{\pm}e^{-{\rm i}\phi_{x}}\\
\sinh\theta_{\mibs k}^{\pm}e^{{\rm i}\phi_{x}}&
-\cosh\theta_{\mibs k}^{\pm}
\end{array}
\right)
\end{equation}
\begin{equation}
V_{\mibs k}^{\pm}=\left(
\begin{array}{cc}
-\sinh\theta_{\mibs k}^{\pm}&
\cosh\theta_{\mibs k}^{\pm}e^{-{\rm i}\phi_{x}}\\
\cosh\theta_{\mibs k}^{\pm}e^{{\rm i}\phi_{x}}&
-\sinh\theta_{\mibs k}^{\pm}
\end{array}
\right)
\end{equation}
\begin{eqnarray}
\cosh\theta_{\mibs k}^{\pm}&=&\frac{1}{\sqrt{2}}\sqrt{\frac{1}
{\epsilon_{\mibs k}^{\pm}}+1},\\
\epsilon_{\mibs k}^{\pm}&=&\sqrt{1-\frac{1}{\mu^{2}}\left(
\frac{j\kappa_{x}\pm \kappa_{y}}{1+j}\right)^{2}},\\
\mu&=&1-\frac{\lambda}{2(J_{x}+J_{y})s},
\end{eqnarray}
the Hamiltonian (\ref{hamihami}) is diagonalized as
\begin{eqnarray}
H=\sum_{\mibs k}\sum_{i=1}^{2}\left[E_{\mibs k}^{+}
\alpha_{\mibs k}^{(i)\dag}
\alpha_{\mibs k}^{(i)}+E_{\mibs k}^{-}
\beta_{\mibs k}^{(i)\dag}\beta_{\mibs k}^{(i)}\right].
\label{eqn:HAB}
\end{eqnarray}
where the energy dispersion $E_{\mibs k}^{\pm}$ is defined by 
$2(J_{x}+J_{y})s\mu\epsilon_{\mibs k}^{\pm}$.
The spin gap $\Delta$ opens at $\mib k=(0, 0)$, and is estimated as
\begin{equation}
\Delta = \sqrt{\lambda^{2}-4(J_{x}+J_{y})s\lambda}.
\label{gapgap}
\end{equation}
We can easily confirm that in the case of $\lambda=0$ this theory 
is identical with the conventional
spin-wave theory and that the lowest mode of the system 
has the linear dispersion. 


In order to formulate thermodynamics, we introduce
the probability, $P_{i\mibs k}^{+}(n_{i}^{+})$ ($P_{i\mibs k}^{-}
(n_{i}^{-})$),  for the state with the wave 
vector $\mib k$ being occupied by 
 $n_{i}^{+}$($n_{i}^{-}$) bosons with
type $\alpha^{(i)}$($\beta^{(i)}$). 
Then the free energy has the form
\begin{eqnarray}
F&=&\sum_{\mibs k,i,n}\left\{
E_{\mibs k}^{+}nP_{i\mibs k}^{+}(n)+ E_{\mibs k}^{-}nP_{i\mibs k}^{-}(n)
\right. \nonumber\\
&-&\left.\left[\mu_{i\mibs k}^{+}P_{i\mibs k}^{+}(n)+
\mu_{i\mibs k}^{-}P_{i\mibs k}^{-}(n)\right]\right\}-TS,
\end{eqnarray}
where $\mu_{i\mibs k}^{\pm}$ is the generalized chemical potential, 
and $S$ is the entropy.
By evaluating 
$P_{i\mibs k}^{\pm}(n)$ which minimizes the free energy $F$, 
we end up with an ordinary formula, 
\begin{eqnarray}
P_{i\mibs k}^{\pm}&=&\left(1-\exp
\left[-E_{\mibs k}^{\pm}/T\right]\right)
\exp\left[-nE_{\mibs k}^{\pm}/T\right].
\end{eqnarray}
Taking the $N\rightarrow\infty$, the resulting constraint
has the form
\begin{eqnarray}
s+\frac{1}{2}
=\frac{1}{\pi^{2}}\int_{0}^{\frac{\pi}{2}}{\rm d}k_{x}
\int_{0}^{\frac{\pi}{2}}{\rm d}k_{y}\left[
\frac{1}{\epsilon_{\mibs k}^{+}}\coth\frac{E_{\mibs k}^{+}}{2T}+
\frac{1}{\epsilon_{\mibs k}^{-}}\coth\frac{E_{\mibs k}^{-}}{2T}\right].
\nonumber\\\label{eqn:consteq}
\end{eqnarray} 
This completes the MSW formulation of our model.
Solving the above equation to determine the chemical potential 
$\lambda$, we then evaluate the spin gap at 
finite temperatures from eq.(\ref{gapgap}).

In the following we discuss the results 
separately for three cases studied in the 
previous section. 

\subsection{Plaquette-structure system}
Let us first show the results obtained for the system with 
plaquette structure. We here focus on the $s=1/2$ system for which 
our NL$\sigma$M approach with saddle-point approximation 
has encountered the problem.
\begin{figure}[htb]
\vspace{0.1cm}
\epsfxsize=8cm 
\centerline{\epsfbox{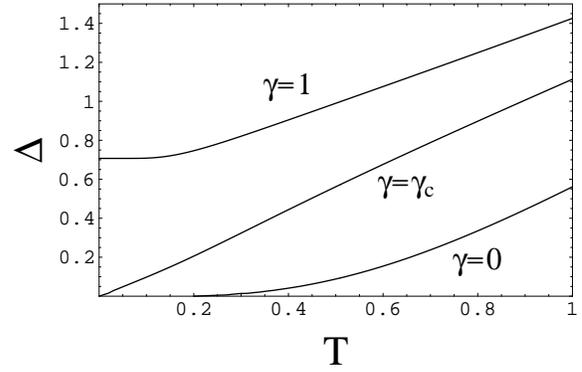}} 
\caption{Spin gap $\Delta$ for 
the plaquette structure calculated within the MSW approximation.
It is shown as a function of the temperature$T$.}
\label{fig:plaT}
\end{figure}
\begin{figure}[htb]
\vspace{0.1cm}
\epsfxsize=8cm 
\centerline{\epsfbox{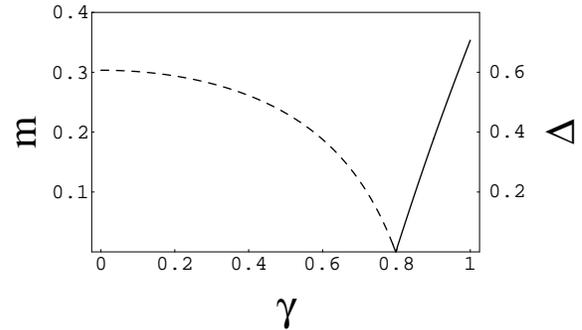}}
\caption{Staggered magnetization $m$ (the dashed line) 
and the spin gap $\Delta$ (the solid line) for the plaquette structure
calculated within the MSW approximation.}
\label{fig:plamag}
\end{figure}
The present MSW approach provides us with not only the temperature 
dependence of the spin gap (Fig. \ref{fig:plaT})
but also the staggered magnetization at $T=0$ (Fig. \ref{fig:plamag}).
We find in Fig. \ref{fig:plaT} that, in contrast with the results 
of the NL$\sigma$M, the solid line for $\gamma = 0$ reproduces 
the gapless behavior at $T=0$. Moreover we can see 
in Fig. \ref{fig:plamag} that when $\gamma=0$, the system has a
finite magnetization ($m= 0.303 $).
The magnetization decreases with increasing $\gamma$ 
and finally vanishes at the critical value $\gamma_{c}=0.798$.
For $\gamma \geq \gamma_{c}$ the plaquette-singlet ground state with 
spin gap is stabilized, as mentioned in the previous section.
These results qualitatively improve the results of 
NL$\sigma$M with saddle-point approximation
since the effect of antiferromagnetic 
correlation may be treated more properly in the MSW approximation. 
Though the MSW can qualitatively describe the correct behavior 
of the quantum phase transition, it may be plausible to evaluate 
various quantities more quantitatively. To this end, we have also 
performed a complementary calculation based on the cluster 
expansion starting from the isolated spin plaquette, up to fourth 
order, which has been developed by several groups 
recently.\cite{cluster}  By applying the Pad\'e approximation 
to the above series expansion, we have confirmed
that there indeed exists 
the quantum phase transition and the critical value $\gamma_{c}$ 
is around 0.3.\cite{Fukumoto}
This largely improves the result of the MSW quantitatively.
The detail for the results of the cluster expansion will be 
reported elsewhere. 

\subsection{Ladder-structure system}
\begin{figure}[htb]
\vspace{0.1cm}
\epsfxsize=8cm 
\centerline{\epsfbox{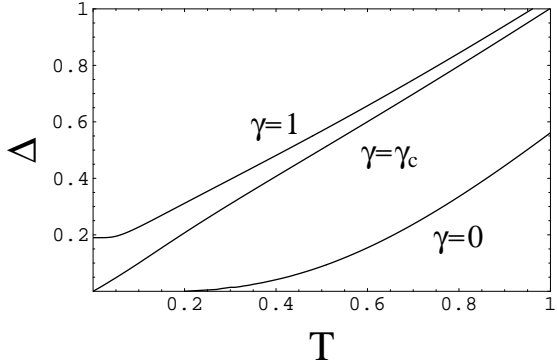}} 
\caption{Spin gap $\Delta$ in the $s=1/2$ ladder structure 
for $j=2$ calculated by the MSW approximation.}
\label{fig:ladT}
\end{figure}

\begin{figure}[htb]
\vspace{0.1cm}
\epsfxsize=8cm 
\centerline{\epsfbox{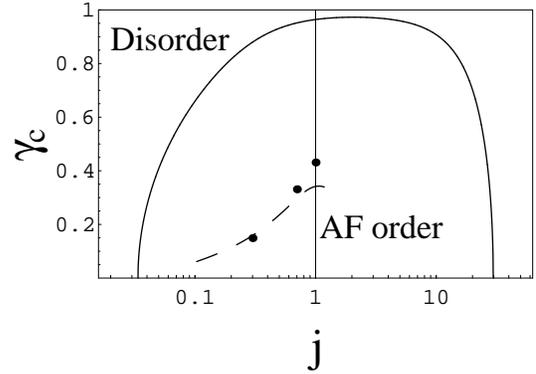}}
\caption{Phase diagram of the ladder system 
calculated by the MSW approximation (see the results 
of ref.\cite{KatoImada}).
The broken line is the phase boundary determined by the 
dimer expansion method up to fifth order combined 
with the Pad\'e approximants. We have also shown the results
of the Quantum Monte Carlo 
calculation\cite{KatoImada} by the dots.}
\label{fig:boundary}
\end{figure}

We next consider the 2D $s=1/2$ system with ladder structure
discussed in the previous section.\cite{KatoImada}
In Fig. \ref{fig:ladT}, the spin gap calculated by the 
MSW is shown as a function of the temperature for several choices
of the bond alternation parameter.
\noindent
It is seen from  Fig. \ref{fig:ladT} that there is a critical 
strength of the bond-alternation parameter $\gamma_c$ at which  
the correlation length (inverse of the spin gap) becomes large
with decreasing temperature, and diverges at zero temperature.
For $\gamma<\gamma_c$, the spontaneous antiferromagnetic order
emerges, but only at zero temperature. In Fig. \ref{fig:boundary}, 
we show the phase diagram in terms of  the 
bond alternation parameter $\gamma$ and the 
anisotropic parameter $j=J_x/J_y$.  If we fix $\gamma$ to be an
appropriate value,  the quantum phase transitions 
may be observed twice with the  increase of $j$. If $j$ is very 
small, the system is approximated by  weekly coupled spin chains with
dimerization. This phase is massive because of the dimer gap for
the spin chain.
When $j$ is increased, the system is driven to the 2D 
antiferromagnetically ordered state. If $j$ is further increased,
the system may be described by a weakly coupled ladder system
which should be again massive, and thus we encounter the second
phase transition. 

\noindent

We should note here that the present 
ladder-structure  system at $T=0$ 
was previously studied by Katoh 
and Imada by MSW theory.\cite{KatoImada}
Our results at $T=0$ indeed coincide with their results.
Furthermore, they have 
performed the Monte Carlo calculations, and confirmed 
the quantum phase transition at the critical strength of the 
bond-alternation parameter, and estimated its value rather 
precisely, which improves the 
results of the MSW quantitatively.  
In this connection, we have also performed
the calculation based on the dimer expansion\cite{cluster}
up to fifth order by taking isolated dimers 
for rungs as a starting point. 
This approach may be complementary to the
Monte Carlo approach of Katoh and Imada.
The phase boundary determined by the dimer expansion method 
with Pad\'e approximation is
plotted in Fig. \ref{fig:boundary} with broken line
for smaller $j$ region, which shows 
fairly good  agreement with Monte Carlo results.

\subsection{Mixed spin system}
\begin{figure}[htb]
\vspace{0.1cm}
\epsfxsize=8cm 
\centerline{\epsfbox{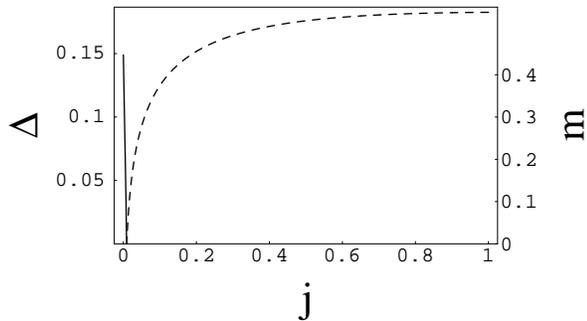}} 
\caption{The staggered magnetization $m$ (the dashed line) and 
the spin gap $\Delta$ (the solid line) for the $s=1/2, 1$ mixed spin  
system calculated by the MSW approximation. The calculated quantities 
are shown as a function $j(=J_{x}/J_{y})$.}
\label{fig:altgap}
\end{figure}

As mentioned in the previous section, the mixed 
spin chains have attracted much current interest. Although we could
describe the quantum phase transition for the $s=1, 3/2$ mixed system
qualitatively by NL$\sigma$M approach, we failed to do
for the $s=1/2, 1$  mixed case, which may be 
most interesting from the viewpoint of experiments.
We address the latter case in this subsection.
In Fig. \ref{fig:altgap}, we show the results for the
spin gap and spontaneous magnetization calculated by MSW
method for mixed spin system with $s=1/2, 1$. 
For $j=0$, the system is reduced to the
mixed spin chain with the periodic array of spins 
as $\cdots 11\frac{1}{2}\frac{1}{2}11\frac{1}{2}\frac{1}{2}\cdots$,
so that it  is in the massive phase because of the 
topological nature of the system.\cite{previous} 
This is consistent with the results of NL$\sigma$M. 
In contrast to NL$\sigma$M approach, however, the spin 
gap rapidly decreases in the vicinity 
of $j=0$, and the systems is driven to the 
ordered phase with a finite staggered magnetization at $j_{c}=0.009$. 

Up to now, the mixed spin chains found experimentally
are known to exhibit the ferrimagnetic ground state.\cite{AltExp} 
It may be expected that the mixed spin chains with
singlet ground state, as discussed here, may be also 
fabricated experimentally in the future. Then such mixed spin chains 
should provide an interesting paradigm of quantum spin systems, 
for which the topological nature of spatial variation of
spins control the low-energy properties of the systems.
In particular, it may be interesting to observe
how such massive systems are driven to the magnetically 
ordered phase in the presence of the interchain coupling, 
impurities, etc.

\section{Summary}
By extending our previous studies on the spin chain 
and ladders, we have systematically investigated 
quantum phase transitions in 2D spin systems with
plaquette and ladder structures.  
In these systems, the lattice structure as well as 
the competing interactions play a crucial role
to determine low-energy properties. 
By computing the spin gap and the spontaneous magnetization, 
by NL$\sigma$M and MSW approaches,
we have confirmed that both of the  above approaches 
describe the quantum phase transition 
qualitatively well, so far as the system with larger spins
 is concerned. 
However, it has turned out that for $s=1/2$ systems (as well as
$s=1/2, 1$ mixed spin systems), the NL$\sigma$M 
with saddle point approximation cannot 
properly describe the phase transition. So, this approach
employed recently by S\'en\'echal may not be appropriate 
to discuss the phase transition even at qualitative level for 
$s=1/2$ spin models. On the other hand, the MSW treatment, which 
may be provide the results comparable to the renormalization group 
treatment of NL$\sigma$M, has been shown to plausibly 
describe the phase transition for $s=1/2$.

In order to give more quantitative discussions
for the quantum phase transitions, we have also shown some preliminary
calculations based on the series expansion methods such as the 
dimer expansion, the plaquette expansion, etc, which 
indeed improve the results of NL$\sigma$M
and MSW. To derive more precise information
on these quantum spin systems, it is desirable to 
extend the cluster expansion to higher orders systematically,
which is now under consideration.

\section{Acknowledgements}
The work is partly supported by a Grant-in-Aid 
>from the Ministry of Education, Science, Sports, and Culture.
%
%
 


\end{document}